\documentclass[aps,pra,preprint,showpacs,amsmath,amssymb]{revtex4}
\input{epsf}

\usepackage{graphicx}
\usepackage{array}
\usepackage{bigstrut}
\usepackage{longtable}
\usepackage{rotating,booktabs}
\usepackage{booktabs,threeparttable}
\usepackage{bm}

\begin{document}

\title{ Dynamic dipole polarizabilities for the low-lying triplet states of helium}

\author{Yong-Hui Zhang$^{1,2}$, Li-Yan Tang$^{2,*}$~\footnotetext{*Email: lytang@wipm.ac.cn},
Xian-Zhou Zhang$^{1}$, and Ting-Yun Shi$^{2}$}

\affiliation {$^{1}$Department of Physics, Henan Normal University,
XinXiang 453007, People's Republic of China}

\affiliation {$^2$State Key Laboratory of Magnetic Resonance and
Atomic and Molecular Physics, Wuhan Institute of Physics and
Mathematics, Chinese Academy of Sciences, Wuhan 430071, People's Republic of China}

\date{\today}

\begin{abstract}
The dynamic dipole polarizabilities for the four lowest triplet states ($2\,^3S$, $3\,^3S$, $2\,^3P$ and $3\,^3P$) of
helium are calculated using the B-spline configuration interaction method.
Present values of the static dipole polarizabilities in the length, velocity and acceleration gauges
are in good agreement with the best Hylleraas results. Also the tune-out wavelengths in the range from 400 nm to 4.2~$\mu$m
for the four lowest triplet states are identified, and the magic wavelengths in the range from 460~nm to 3.5~$\mu$m for the $2\,^3S \to 3\,^3S$, $2\,^3S \to 2\,^3P$,
and $2\,^3S \to 3\,^3P$ transitions are determined. We show that the tune-out wavelength of $2\,^3S$ state is 413.038 28(3) nm, which corroborates
the value of Mitroy and Tang (Phys. Rev. A 88, 052515 (2013)), and the magic wavelength around 1066 nm for the $2\,^3S \to 3\,^3P$ transition can be
expected for precision measurement to determine the ratio of transition matrix elements $(2\,^3S \to 2\,^3P) / (3\,^3P\to 6\,^3S)$.
\end{abstract}

 \maketitle

\section{introduction}

Precise calculations of dynamic dipole polarizabilities for atoms are of interest
due to its importance in a number of applications. First, dynamic dipole polarizabilities can
be used directly to analyse the ac Stark shift to pursue higher-precision atomic clocks~\cite{mitroy10a,tang13b}.
Second, investigation of the dynamic dipole polarizabilities can derive the magic wavelengths and tune-out wavelengths,
which open a new route to determine the line strength ratio~\cite{tang13b,herold12a} and to test the relativistic and
quantum electrodynamic (QED) effects upon the transition matrix element not on the energy~\cite{mitroy13b,henson15}. And at last, since
both of the trapping potential depth and the photon-scattering rate are dependent on the polarizabilities, the calculations of the
dynamic dipole polarizabilities can provide reliable reference for experimental design to trap atoms in efficiency~\cite{safronova12c, notermans14}.

As the simplest two-electron system, the accurate theoretical calculations and experimental
measurements of the energy levels for helium can be used to test the the three-body bound QED theory~\cite{drake08, eyler08}, to determine the fine structure constant with high-precision~\cite{lewis78,pachucki10,smiciklas10}, to extract the
nuclear information without resorting to any model~\cite{rooij11a,cancio12a}, and to develop the multi-electron atomic structure theory~\cite{eyler08, yerokhin10a}.
Recently, the resonance transition $2\,^3S \to 2\,^3P$ and the doubly-forbidden transition $2\,^3S \to 2\,^1S$ of
helium isotopes have attracted great interest for the determination of nuclear charge radius
difference~\cite{cancio12a,rooij11a, leeuwen06, notermans14}. Combined the laser cooling with magneto-optical trap techniques, the transitions
$2\,^3S \to 2\,^3P$ and $2\,^3S \to 3\,^3P$ of helium are also demonstrated to produce high density quantum gas~\cite{tychkov04}.
The key point to improve the experimental measurement precision for helium is setting the laser frequency at the magic wavelength to
eliminate effectively the ac Stark shift induced by the trap light.

At present, there are lots of literatures focused on the accurate calculations of the energy and polarizabilities~\cite{drake99a,drake01,drake02a,tang13,jamieson95,masili03,kar12} for the ground state of helium.
For example, the non-relativistic ground-state energy has been achieved up to 46 digits~\cite{schwartz06b},
and the static dipole polarizability of the ground-state helium, which includes the effect of mass polarization, the
relativistic and leading QED corrections, has been determined to 1.383 191(2) within 2 ppm accuracy~\cite{pachucki00a}.
However, compared with the ground state, there are very few calculations
of dynamic polarizabilities for the triplet states of helium. As we known, for the metastable state $2\,^3S$ of helium, Glover {\em et al.} listed the rigorous
upper and lower bounds of the dynamic dipole polarizabilities~\cite{glover77a}. Chung provided dynamic polarizabilities for
frequencies up to the second excitation threshold by using a variation-perturbation scheme~\cite{chung77a}. Chen used a
configuration interaction (CI) scheme with B-spline functions~\cite{boor78} to improve the convergence of the dynamic
dipole polarizabilities~\cite{chen95,chen95a}. And R$\acute{e}$rat~{\em et al.}
presented the dynamic dipole polarizabilities of helium at both real and imaginary
frequencies using time-dependent gauge-invariant method~\cite{rerat93a}. In 2005, Chernov {\em et al.}
calculated the dynamic polarizabilities~\cite{chernov05a} by using the quantum defect Green function formalism.
For others triplet states of helium, there are fewer reports can be referenced~\cite{rerat94,chernov05a}.

In this work, firstly, we have performed the calculations of static dipole polarizabilities for the low-lying triplet states
$2\,^3S$, $2\,^3P$, $3\,^3S$ and $3\,^3P$ of helium with the configuration
interaction method based on B-spline functions in the length, velocity and acceleration
gauges. Then the dynamic dipole polarizabilities of
$2\,^3S$, $2\,^3P$ for frequencies below the second excitation threshold, and $3\,^3S$, $3\,^3P$ for
frequencies below the first ionization threshold are calculated utilizing
oscillator strengths and energy differences obtained in the length gauge. In addition, using the dynamic dipole polarizabilities,
the magic wavelengths for the three transitions $2\,^3S \to 3\,^3S$, $2\,^3S \to 2\,^3P$, and $2\,^3S \to 3\,^3P$, and the tune-out wavelengths
for the four lowest triplet states $2\,^3S$, $2\,^3P$, $3\,^3S$, and $3\,^3P$ are determined with high accuracy.

\section{dipole polarizability}

The dynamic dipole polarizability of the magnetic sub-level $|L_gM_g\rangle$ is
\begin{equation}
\alpha_{L_gM_g}(\omega)=\alpha_1(\omega)+\dfrac{3M_g^2-L_g(L_g+1)}{L_g(2L_g-1)}\alpha_1^T(\omega)
\, ,\label{e1}
\end{equation}
where $\alpha_1(\omega)$ and $\alpha_1^T(\omega)$ are the dynamic scalar and tensor dipole polarizabilities respectively,
they are expressed as the summation of all allowed-transition intermediate states, including the continuum,
\begin{eqnarray}
\alpha_1(\omega)=\sum_{n\neq g}\dfrac{f_{gn}^{(1)}}{(\Delta E_{gn})^2-\omega^2}
\, ,\label{e2}
\end{eqnarray}
\begin{eqnarray}
\alpha_1^T(\omega)=\sum_{n\neq g}
(-1)^{L_g+L_n}\sqrt{\dfrac{30(2L_g+1)L_g(2L_g-1)}{(2L_g+3)(L_g+1)}}
\left \{
\begin{array}{ccc}
1   &1   &2\\
L_g &L_g &L_n\\
\end{array}
\right \}
\dfrac{f_{gn}^{(1)}}{(\Delta E_{gn})^2-\omega^2}
\, .\label{e3}
\end{eqnarray}
In the above formula, $\Delta E_{gn}$ is the transition energy between the initial state $|N_gL_gM_g\rangle$ and the intermediate state $|N_nL_nM_n\rangle$,
$\omega$ is the photon energy of external electric field, and the dipole oscillator strength $f_{gn}^{(1)}$ have
different expressions in the length, velocity, and acceleration gauges respectively,
\begin{equation}
f_{gn}^{(1)}=\dfrac{2|\langle N_gL_g\|
\sum\limits_{i=1, 2}r_iC^{(1)}(\hat{r}_i)
                                \|N_nL_n\rangle|^2\Delta E_{gn}}{3(2L_g+1)}
\, , \label{e4}
\end{equation}
\begin{equation}
f_{gn}^{(1)}=\dfrac{2|\langle N_gL_g\|
\sum\limits_{i=1, 2}\dfrac{d}{dr_i}C^{(1)}(\hat{r}_i)
+b(\ell_k;\ell_{\ell})r_i^{-1}C^{(1)}(\hat{r}_i)
                                \|N_nL_n\rangle|^2(\Delta E_{gn})^{-1}}{3(2L_g+1)}
\, , \label{e5}
\end{equation}
\begin{equation}
f_{gn}^{(1)}=\dfrac{2|\langle N_gL_g\|
\sum\limits_{i=1, 2}Zr_i^{-2}C^{(1)}(\hat{r}_i)
                                \| N_nL_n\rangle|^2(\Delta E_{gn})^{-3}}{3(2L_g+1)}
\, , \label{e6}
\end{equation}
where $\sum\limits_{i=1, 2}r_iC^{(1)}(\hat{r}_i)$ being the electronic dipole transition operator of two-electron system,
$Z$ is the nuclear charge number, $\ell_k$ or $\ell_{\ell}$ are the orbital quantum number of a electron,
and the function $b(\ell;\lambda)$ is defined as
\begin{eqnarray}
\left\{
\begin{array}{lcl}
b(\ell;\ell-1) &=& \ell+1 \\
b(\ell;\ell+1) &=& -\ell \,.
 \label{e7}
\end{array}
\right.
\end{eqnarray}

According to the Eqs.~(\ref{e2}) and ~(\ref{e3}), for the case of $L_g=0$ initial state,
the dynamic scalar and tensor dipole polarizabilities are
\begin{eqnarray}
\alpha_1(\omega)&=&\alpha_1(P, \omega)\\
\alpha_1^T(\omega)&=&0
\, ,\label{e8}
\end{eqnarray}
where $\alpha_1(P, \omega)$ represents the contributions of the intermediate state with the angular momentum number $L_n=1$.

For the initial state of $L_g=1$, the dynamic scalar and tensor dipole polarizabilities are expressed as
\begin{eqnarray}
\alpha_1(\omega)&=&\alpha_1(S, \omega)+\alpha_1(P, \omega)+\alpha_1(D, \omega) \\
\alpha_1^T(\omega)&=&-\alpha_1(S, \omega)+\dfrac{1}{2}\alpha_1(P, \omega)-\dfrac{1}{10}\alpha_1(D, \omega)
\, ,\label{e9}
\end{eqnarray}
where $\alpha_1(S, \omega)$ and $\alpha_1(D, \omega)$ are the contributions of the natural parity state
$(ss)S$ and $(sd)D$ respectively, and $\alpha_1(P, \omega)$ is the contribution of the unnatural parity
state of $(pp')P$ electron configuration.

In order to calculate the dynamic dipole polarizabilities, the fundamental atomic structure information of energies and wavefunctions are needed to obtain firstly.
In our calculations, the configuration interaction method based on B-spline functions are adopted to
get the energies and wavefunctions for helium.

\section{configuration interaction with B-spline basis}

The Hamiltonian for two-electron system is given in second-quantized
form as
\begin{equation}
H=\sum\limits_i\varepsilon_ia_i^+a_i+\dfrac{1}{2}\sum\limits_{ijk\ell}g_{ijk\ell}
a_i^+a_j^+a_{\ell}a_k
\, , \label{e10}
\end{equation}
where $\varepsilon_i$ is the $i$th energy eigenvalue of the single-particle
Schr$\ddot{o}$dinger equation, $g_{ijkl}$ is two-particle matrix element of the Coulomb interaction, and $a_i^+$
and $a_i$ are creation and annihilation operators for the $i$th electron respectively. The single-particle quantum state
is presented as  $|n_i \ell_i m_i m_{s_i}\rangle$, here $n_i$ is the principal quantum number, $\ell_i$
is the orbital angular momentum, $m_i$ and $m_{s_i}$ are the orbital and
spin angular momentum projection, respectively.

The two-electron wavefunction $\psi_{ij}(LS)$ is expressed
as a linear combination of
configuration-state wavefunctions $\phi_{ij}(LS)$,
\begin{equation}
\psi_{ij}(LS)=\sum_{ij}c_{ij}\phi_{ij}(LS)
\, , \label{e11}
\end{equation}
and the configuration-state wavefunction has the following expression,
\begin{equation}
\phi_{ij}(LS)=\eta_{ij}\sum\limits_{m_im_j}\sum\limits_{m_{s_i}m_{s_j}}
\langle\ell_im_i;\ell_jm_j|LM\rangle\langle1/2m_{s_i};1/2m_{s_j}|SM_S\rangle
a_i^+a_j^+|0\rangle
\, , \label{e12}
\end{equation}
where $\eta_{ij}$ is a normalization constant given by
\begin{eqnarray}
\eta_{ij}=\left \{
\begin{array}{cc}
1,                   &i\neq j \\
\frac{1}{\sqrt{2}}, &i=j \,.
\end{array}
\right.
\label{e13}
\end{eqnarray}
The Clebsch-Gordan coefficients $\langle\ell_im_i;\ell_jm_j|LM\rangle$ and
$\langle1/2m_{s_i};1/2m_{s_j}|SM_S\rangle$ represent $\ell\ell$ and
$ss$ coupling, respectively, $|0\rangle$ is the vacuum state and $a_i^+|0\rangle$
represents the $i$th eigen-wavefunction of the single-particle Schr$\ddot{o}$dinger equation
with energy eigenvalue $\varepsilon_i$. The configuration-state wavefunctions are independent
of magnetic quantum numbers of $m_i$, $m_j$, $m_{s_i}$ and $m_{s_j}$. From the interchange
symmetery of the Clebsch-Gordan coefficients, it follows that
\begin{equation}
\phi_{ji}(LM)=(-1)^{\ell_i+\ell_j+L+S}\phi_{ij}(LM)
\, ,\label{e14}
\end{equation}
which implies $\phi_{ii}(LM)=0$ unless $L+S$ is even.

According to the expansion of the wavefunctions, the matrix elements of Hamiltonian
is
\begin{equation}
\langle\psi_{ij}(LM)|H|\psi_{k\ell}(LM)\rangle=
\sum_{k\ell}(\varepsilon_i+\varepsilon_j)c_{ij}c_{k\ell}\delta_{ik}\delta_{j\ell}+\sum_{ij,k\ell}V_{ij,k\ell}
c_{ij}c_{k\ell}
\, ,\label{e16}
\end{equation}
where the potential energy matrix element $V_{ij,k\ell}$ between different configurations is
\begin{eqnarray}
V_{ijk\ell}=\eta_{ij}\eta_{kl}
&&\left[
\sum_{\nu}(-1)^{\ell_j-\ell_k+L+\nu}
\left \{
\begin{array}{ccc}
\ell_i      &\ell_j &L   \\
\ell_{\ell} &\ell_k &\nu \\
\end{array}
\right \}
X_{\nu}(ijk\ell)+   \right.
\\ \nonumber
&&\left.
\sum_{\nu}(-1)^{\ell_j-\ell_k+S+\nu}
\left \{
\begin{array}{ccc}
\ell_i  &\ell_j      &L   \\
\ell_k  &\ell_{\ell} &\nu \\
\end{array}
\right \}
X_{\nu}(ij\ell k)
\right ]\, .\label{e17}
\end{eqnarray}
The quantity $X_{\nu}(ijk\ell)$ in the above equation is given by
\begin{equation}
X_{\nu}(ijk\ell)=(-1)^{\nu}\langle\ell_i\|C^{\nu}\|\ell_k\rangle
                           \langle\ell_j\|C^{\nu}\|\ell_{\ell}\rangle
R_{\nu}(ijk\ell)
\, ,\label{e18}
\end{equation}
where $\langle\ell_i\|C^{\nu}\|\ell_k\rangle$ is angular
reduced matrix element,
\begin{eqnarray}
\langle\ell_i\|C^{\nu}\|\ell_k\rangle=(-1)^{\ell_i}\sqrt{(2\ell_i+1)(2\ell_k+1)}
\left \{
\begin{array}{ccc}
\ell_i  &\nu &\ell_k\\
0       &0   &0    \\
\end{array}
\right \}
\, .\label{e19}
\end{eqnarray}
The two-electron radial integral $R_{\nu}(ijk\ell)$ of the Coulomb
interaction is written as
\begin{equation}
R_{\nu}(ijk\ell)=\int\int r_1^2dr_1 r_2^2dr_2 R_i(r_1)R_k(r_1)
\dfrac{r_<^{\nu}}{r_>^{\nu+1}}
R_j(r_2)R_{\ell}(r_2)
\, ,\label{e20}
\end{equation}
where $r_<$ and $r_>$ are the minimum and maximum of $r_1$ and $r_2$,
and $R_i(r)$ is the radial wavefunction of the $i$th single-electron orbital.

Using the variational method, the followed configuration interaction equations can be obtained,
\begin{equation}
\sum_{k\ell}\big[(\varepsilon_i+\varepsilon_j)\delta_{ik}\delta_{jl}+V_{ijkl}\big]c_{kl}=\lambda c_{ij}
\, ,\label{e22}
\end{equation}
where $\lambda$ and ${c_{ij}}$ are the eigen-energy and  eigen-wavefunction for two-electron atoms, respectively.

Before solved the CI equations, the energies and wavefunctions for single-electron
orbital are obtained firstly. In our calculations, B-splines are used to expand the radial wavefunction
for the $i$th single-electron orbital,
\begin{equation}
R_i(r)=\sum_jc_j^iB_j(r)
\, .\label{e23}
\end{equation}
where $\{c_j^i\}$ are the expansion coefficients, and the following exponential knots are employed,
\begin{eqnarray}
\left \{
\begin{array}{ll}
t_i=0, &i=1,2,\cdot\cdot\cdot,k-1;\\
t_{i+k-1}=R_0\dfrac{\exp\left[\gamma R_0\left(\dfrac{i-1}{N-2}\right)\right]-1}{\exp[\gamma R_0]-1}, &i=1,2,\cdot\cdot\cdot,N-1;\\
t_i=R_0, &i=N+k-1,N+k\,.\\
\end{array}
\right.
\label{e24}
\end{eqnarray}
where $R_0$ is the box size, which need to be chosen large enough to make sure the contributions to dynamic dipole polarizabilities from
higher excited-state are included, especially when the photon energy $\omega$ is large.
The non-linear parameter $\gamma$ is also need to be adjusted to get more accurate ground-state energy of helium,
then the value of $\gamma$ is fixed the same for all the triplet states to simplify the integral of B-splines.

\section{results and discussions}

In our calculation, $R_0=200$ a.u. and $\gamma=R_0\times0.038$ are used throughout the paper.
Using the fixed values of $R_0$ and $\gamma$, we get the ground-state energy $-2.879~0284$ a.u. under S-wave approximation with 30 B-splines
of order 7, which has 7 significant digits with the S-wave limit value $-2.879~028~767~29$ a.u.~\cite{decleva95a}.

\subsection{Energies and Oscillator Strengths }

%=====================================================================================================================
\begin{table*}\scriptsize
\caption{\label{energy1}Convergence of the energies (in a.u.) for the metastable state $2\,^3S$ and the oscillator strengths $f_{2\,^3S\to2\,^3P}^{(L)}$ in the length
gauge of helium as the number of B-splines $N$ and partial waves $\ell_{max}$ increased. The numbers in parentheses of the extrapolated values give the computational uncertainties.}
\begin{ruledtabular}
\begin{tabular}{lllllll}
 \multicolumn{1}{c}{}
&\multicolumn{3}{c}{Energy}
&\multicolumn{3}{c}{$f_{2\,^3\,S \to 2\,^3P}^{(L)}$}
\\
\hline
 \multicolumn{1}{c}{$\ell_{max}$}
&\multicolumn{1}{c}{N=30} &\multicolumn{1}{c}{N=35} &\multicolumn{1}{c}{N=40}
&\multicolumn{1}{c}{N=30} &\multicolumn{1}{c}{N=35} &\multicolumn{1}{c}{N=40}
\\
 \cline{2-4} \cline{5-7}
2 &$-$2.175 220 4147  &$-$2.175 220 4306  &$-$2.175 220 4345   &0.539 818 2380   &0.539 818 2108   &0.539 818 2056 \\
3 &$-$2.175 227 0950  &$-$2.175 227 1164  &$-$2.175 227 1220   &0.539 204 6079   &0.539 204 5519   &0.539 204 5388 \\
4 &$-$2.175 228 5828  &$-$2.175 228 6093  &$-$2.175 228 6165   &0.539 117 1546   &0.539 117 0681   &0.539 117 0460  \\
5 &$-$2.175 229 0255  &$-$2.175 229 0559  &$-$2.175 229 0647   &0.539 096 9772   &0.539 096 8626   &0.539 096 8313  \\
6 &$-$2.175 229 1847  &$-$2.175 229 2183  &$-$2.175 229 2283   &0.539 090 7709   &0.539 090 6327   &0.539 090 5931  \\
7 &$-$2.175 229 2501  &$-$2.175 229 2859  &$-$2.175 229 2970   &0.539 088 4620   &0.539 088 3052   &0.539 088 2583 \\
8 &$-$2.175 229 2796  &$-$2.175 229 3171  &$-$2.175 229 3289   &0.539 087 4793   &0.539 087 3088   &0.539 087 2559 \\
9 &$-$2.175 229 2939  &$-$2.175 229 3326  &$-$2.175 229 3450   &0.539 087 0173   &0.539 086 8365   &0.539 086 7790  \\
10&$-$2.175 229 3013  &$-$2.175 229 3407  &$-$2.175 229 3536   &0.539 086 7826   &0.539 086 5947   &0.539 086 5336 \\
Extrap. & \multicolumn{3}{c}{$-$2.175 229 36(2)} & \multicolumn{3}{c}{0.539 086 4(3)}    \\
\end{tabular}
\end{ruledtabular}
\end{table*}

Table~\ref{energy1} is the convergence of the energies for the metastable state $2\,^3S$ and the oscillator strengths of $2\,^3S \to 2\,^3P$ transition
in the length gauge of helium as the number of basis set and partial waves increased.
For the energy, increase of the number of partial wave change less than the number of B-spline N increased. This convergent style for the energy suggested that we can fix partial wave (in our work we fix $\ell_{max} = 10$), then increase the number of B-spline N to avoid too enormous number of CI. Considering both the effect from N and $\ell_{max}$, the extrapolated values are given in the last line of the Table~\ref{energy1}. The final converged value for the energy is $-$2.175 229 36(2) a.u., which is in excellent agreement with the result $-$2.175 229 378 176 a.u. of Cann and Thakkar~\cite{cann92}.  The extrapolated oscillator strength 0.539 0864(3) has 6 significant digits with the value 0.539~0861 of Drake~\cite{drake96a}.

%=======================================================================================================
\begin{table*}
\caption{\label{energy2}Comparison of the energies (in a.u.) in the lengthy gauge for the four lowest triplet states of helium.
The numbers in the parentheses are the computational uncertainties. }
\begin{ruledtabular}
\begin{tabular}{lllll}
 \multicolumn{1}{c}{State}
&\multicolumn{1}{c}{Present}             &\multicolumn{1}{c}{Ref.~\cite{chen94}}
&\multicolumn{1}{c}{Ref.~\cite{cann92}}  &\multicolumn{1}{c}{Ref.~\cite{drake96a}}\\
\hline
$2\,^3S$   &$-$2.175 229 36(2) &$-$2.175 228 8 &$-$2.175 229 378 176  &$-$2.175 229 378 236 791 30 \\
$3\,^3S$   &$-$2.068 689 07(2) &$-$2.068 688 8 &$-$2.068 689 067 283  &$-$2.068 689 067 472 457 19 \\
$2\,^3P$   &$-$2.133 164 17(2) &$-$2.133 163 4 &$-$2.133 164 181 6    &$-$2.133 164 190 779 273(5) \\
$3\,^3P$   &$-$2.058 081 08(2) &$-$2.058 080 6 &$-$2.058 081 077 2    &$-$2.058 081 084 274 28(4)  \\
\end{tabular}
\end{ruledtabular}
\end{table*}

A similar convergence pattern exists for the energies and oscillator strengths in the velocity and
acceleration gauges for the other triplet states of helium.
The final convergent results of the energies in the lengthy gauge are presented in Table~\ref{energy2}.
Our energies are much more accurate than the values~\cite{chen94} by two order of magnitudes,
which are obtained by using the B-splines CI method with different number of configuration states.
And our results for the $2\,^3S$ and $2\,^3P$ states have 8 significant digits with the explicitly correlated calculations~\cite{cann92} and the Hylleraas results~\cite{drake96a}.

%========================================================================================
\begingroup
\squeezetable
\begin{table*}
\caption{\label{oscillator}Comparison of the oscillator strengths in three different gauges for helium.
The numbers in the parentheses are the computational uncertainties. The values of Ref.~\cite{cann92} is the average of the length and velocity gauges, the results of Ref.~\cite{drake96a} are in the length gauge, and Ref.~\cite{alexander06} gives the values in three different gauges.}
\begin{ruledtabular}
\begin{tabular}{lllllllll}
&  \multicolumn{3}{c}{Present}  &\multicolumn{1}{c}{Ref.~\cite{cann92}} & \multicolumn{1}{c}{Ref.~\cite{drake96a}}
& \multicolumn{3}{c}{Ref.~\cite{alexander06}} \\
 \cline{2-4} \cline{5-5} \cline{6-6}\cline{7-9}
 \multicolumn{1}{c}{Transition}
&\multicolumn{1}{c}{$f_{gn}^{(L)}$}   &\multicolumn{1}{c}{$f_{gn}^{(V)}$}    &\multicolumn{1}{c}{$f_{gn}^{(A)}$}
& & & \multicolumn{1}{c}{$f_{gn}^{(L)}$}   &\multicolumn{1}{c}{$f_{gn}^{(V)}$}    &\multicolumn{1}{c}{$f_{gn}^{(A)}$} \\
$2\,^3S \to 2\,^3P$    &0.539 0864(3)   &0.539 0865(2)    &0.539 078(6)    &0.5391   &0.539 0861 &0.5392(8) &0.539(3) &0.56(3)  \\
$3\,^3S \to 3\,^3P$    &0.890 8518(2)   &0.890 8518(4)    &0.890 83(3)     &0.8910   &0.890 8513 &0.890(2)  &0.889(7) &0.85(6)  \\
$2\,^3P \to 3\,^3D$    &0.610 2255(2)   &0.610 2255(2)    &0.610 2247(3)   &0.610 24 &0.610 2252  &0.611(2)  &0.609(2) &0.609(3) \\
$2\,^3P \to 2\,^3P^e$  &0.180 480 28(2)  &0.180 4803(2)    &0.180 4803(3)   &         \\
$3\,^3P \to 4\,^3D$    &0.477 5943(2)   &0.477 5943(2)    &0.477 593(3)    &0.477 60 &0.477 5938  &0.474(3) &0.476(1) &0.494(5) \\
$3\,^3P \to 3\,^3P^e$  &0.135 420 99(2)  &0.135 420 99(3) &0.135 420 98(4)      &         \\
\end{tabular}
\end{ruledtabular}
\end{table*}
\endgroup

Table~\ref{oscillator} lists the comparison of the oscillator strengths for some selected transitions. The
superscripts $(L)$, $(V)$, and $(A)$ represent results obtained in the length,
velocity and acceleration gauges, respectively. For the dipole oscillator strength of $2\,^3S \to 2\,^3P$ transition,
the value in the acceleration gauge is less accurate than the results from length and velocity gauges, but our results for $2\,^3S \to 2\,^3P$ in three gauges are correspondingly much more accurate than the values in different gauges of Ref.~\cite{alexander06} by three order of magnitudes. All of our results in Table~\ref{oscillator}
are much more accurate than the previous values of Refs.~\cite{cann92,alexander06}. And for the oscillator strengths of other transitions, our results in the length and velocity gauge are in excellent agreement with the Hylleraas calculations of Drake~\cite{drake96a}. In addition, the oscillator strengths from the initial states $2\,^3P$ and $3\,^3P$ transit to the unnatural parity states $2\,^3P^e$ and $3\,^3P^e$ are also listed in the Table~\ref{oscillator}.

\subsection{Static Dipole Polarizabilities }

%=====================================================================================================================
\begingroup
\squeezetable
\begin{table*}
\caption{\label{static}Convergence of the static dipole polarizabilities (in a.u.) in three different gauge for the metastable state $2\,^3S$ of helium as the number of B-splines $N$ and partial waves
$\ell_{max}$ increased. The numbers in parentheses give the computational uncertainties.}
\begin{ruledtabular}
\begin{tabular}{llllllllll}
 \multicolumn{1}{c}{}
&\multicolumn{3}{c}{$\alpha_1^{(L)}(0)$}
&\multicolumn{3}{c}{$\alpha_1^{(V)}(0)$}
&\multicolumn{3}{c}{$\alpha_1^{(A)}(0)$}   \\
 \hline
 \multicolumn{1}{c}{$\ell_{max}$}
&\multicolumn{1}{c}{N=30} &\multicolumn{1}{c}{N=35} &\multicolumn{1}{c}{N=40}
&\multicolumn{1}{c}{N=30} &\multicolumn{1}{c}{N=35} &\multicolumn{1}{c}{N=40}
&\multicolumn{1}{c}{N=30} &\multicolumn{1}{c}{N=35} &\multicolumn{1}{c}{N=40}\\
 \cline{2-4} \cline{5-7} \cline{8-10}
2&315.433 490 &315.433 397 &315.433 373 &315.171 086 &315.170 980  &315.170 949 &312.742 586 &312.745 210 &312.745 924\\
3&315.606 026 &315.605 905 &315.605 872 &315.571 449 &315.571 315  &315.571 274 &315.072 959 &315.076 939 &315.078 064\\
4&315.626 281 &315.626 136 &315.626 095 &315.618 705 &315.618 550  &315.618 502 &315.463 517 &315.468 762 &315.470 301\\
5&315.630 187 &315.630 024 &315.629 976 &315.627 938 &315.627 769  &315.627 715 &315.562 445 &315.568 761 &315.570 684\\
6&315.631 213 &315.631 037 &315.630 984 &315.630 396 &315.630 219  &315.630 160 &315.594 442 &315.601 601 &315.603 855\\
7&315.631 548 &315.631 364 &315.631 306 &315.631 205 &315.631 023  &315.630 960 &315.606 534 &315.614 322 &315.616 846\\
8&315.631 678 &315.631 487 &315.631 426 &315.631 515 &315.631 330  &315.631 265 &315.611 603 &315.619 840 &315.622 577\\
9&315.631 735 &315.631 540 &315.631 476 &315.631 649 &315.631 462  &315.631 396 &315.613 883 &315.622 434 &315.625 333\\
10&315.631 763 &315.631 565 &315.631 500 &315.631 712 &315.631 524  &315.631 457 &315.614 962 &315.623 727 &315.626 746\\
Extrap. &\multicolumn{3}{c}{315.631 5(2)} &\multicolumn{3}{c}{315.631 4(2)} &\multicolumn{3}{c}{315.63(2)}\\
\end{tabular}
\end{ruledtabular}
\end{table*}
\endgroup

Table~\ref{static} gives the convergence of the static dipole polarizabilities for the metastable state $2\,^3S$ of helium
as the number of basis set and partial waves increased, and the last line lists the extrapolated values.
From this table, we can see in the length and velocity gauges, the convergence style are the same, the results are decreased as the number of basis sets N increased for a same $\ell_{max}$. However in the acceleration gauge, the values are increased as the number of basis sets N increased for a same $\ell_{max}$. The final convergent value in the length gauge is 315.631 5(2), which has 6 significant digits compared with the most accurate Hylleraas value 315.631 47(1) of Yan~\cite{yan00b}.

%=================================================================================================
\begin{table*}
\caption{\label{static2}Comparison of the static dipole
polarizabilities (in a.u.) for helium. The numbers in parentheses give the computational uncertainties.}
\begin{ruledtabular}
\begin{tabular}{lllll}
 \multicolumn{1}{c}{state}
&\multicolumn{1}{c}{$\alpha_1^{(L)}(0)$}          &\multicolumn{1}{c}{$\alpha_1^{(V)}(0)$}
&\multicolumn{1}{c}{$\alpha_1^{(A)}(0)$}          &\multicolumn{1}{c}{Ref.~\cite{yan00b}} \\
$2\,^3S$  &315.6315(2)   &315.6314(2)   &315.63(2)    &315.631 47(1) \\
$3\,^3S$  &7937.584(2)   &7937.583(2)   &7937.4(2)    &7937.58(1)    \\
$2\,^3P$  &46.70793(4)   &46.70794(4)   &46.71(2)    &46.707 7482(3)\\
$3\,^3P$  &17305.67(3)   &17305.67(4)   &17311(2)    &17305.598(3)  \\
\end{tabular}
\end{ruledtabular}
\end{table*}

Table~\ref{static2} gives the comparison of static dipole polarizabilities for the four lowest triplet states of helium.
The results between the length and velocity gauges are in perfect agreement.
The values obtained in the acceleration gauge are less accurate than the results of lengthy and velocity gauges by two order of magnitudes. Present results for the $2\,^3S$ and $3\,^3S$ states in the length and velocity gauges agree with the Hylleraas values~\cite{yan00b} at the $10^{-7}$ level, and our values for the $2\,^3P$ and $3\,^3P$ states in the length and velocity gauges agree with the Hylleraas values~\cite{yan00b} at the $10^{-6}$ level. For the acceleration gauge, present $\alpha_1^{(A)}(0)$ for $2\,^3P$ and $3\,^3P$ states just have 3 singificant digits compared with the Hylleraas values~\cite{yan00b}.

\subsection{Dynamic Dipole Polarizabilities}

Table~\ref{dynamic} lists the dynamic dipole polarizabilitity for the metastable state $2\,^3S$ of helium
for some selective frequency from 0 to 0.12 a.u., the figures in parentheses represent computational uncertainties.
It seen clearly from this table, all of our values have at least 5 significant digits except the
results of $\omega=0.04$ a.u., $\omega=0.110$ a.u., and $\omega=0.115$ a.u., which only have 4 significant digits.
That's because there is always a tune-out wavelength located in the vicinity of these positions~\cite{mitroy13b},
the relativistic and finite nuclear mass corrections may effect the uncertainities of dynamic dipole polarizabilities.

Table~\ref{dynamic} also makes a comparison of the present results with available values from other
literatures~\cite{chen95a,rerat93a,chung77a}. All of ours results lie within the boundary of Glover and
Weinhold's~\cite{glover77a}, which gives
the rigorous upper and lower limits for the dynamic dipole polarizability at a wide frequency range.
In the low-frequency region, our values are in good agreement with Ref.~\cite{chen95a}, which are also obtained by using B-spline CI method.
For example, ours values have the same five significant digits as theirs. As the frequency $\omega$ increased, the differences
between present results and values of Ref.~\cite{chen95a} increased, especially for the $\omega=0.115$ a.u.,
the difference of the dynamic polarizabilities can reach to about 2.2 $a_0^3$. The reason for
this is that the box size $R_0$ of B-spline adopted in present and Ref.~\cite{chen95a} calculations are different.
As $\omega$ increased, $R_0$ should be chosen big enough to
make sure the transition to high-excited states, especially the transition energies of those excited states near $\omega$,
can be included in the calculation of polarizabilities.
For example, if the box size $R_0=200$ a.u. is adopted, we get $\alpha_1(0.140)=1.910 12(92)$ $a_0^3$
and $\alpha_1(0.145)=-65.327(5)$ $a_0^3$. If the box size is set as $R_0=50$ a.u., then
we get $\alpha_1(0.140)=1.57(6)$ $a_0^3$ and $\alpha_1(0.145)=-68.6(6)$ $a_0^3$, which are less accurate than the values from $R_0=200$ a.u..
However, the box size is not the bigger the better for the B-spline CI calculation, oppositely,
the loss of accuracy will occur under the same number of B-spline for the big box size. In order to get more accurate values, the number of B-spline should be increased,
which makes the number of CI increased exponentially and slows down the convergent process of our calculations. So in our practical calculation, we
need to chose appropriate $R_0$ to get accurate value for large $\omega$ and to avoid large number of CI at the same time.

%===========================================================================================================================
\begin{table*}
\caption{\label{dynamic}Comparison of the dynamic dipole polarizabilities (in a.u.) of the
$2\,^3S$ state for the He atom. The numbers in parentheses give the computational uncertainties. }
\begin{ruledtabular}
\begin{tabular}{cccccc}
 \multicolumn{1}{c}{$\omega$}
&\multicolumn{1}{c}{Present}   &\multicolumn{1}{c}{Ref.\cite{glover77a}}         &\multicolumn{1}{c}{Ref.\cite{chung77a}}
&\multicolumn{1}{c}{Ref.\cite{chen95a}}
&\multicolumn{1}{c}{Ref.\cite{rerat93a}} \\
\hline
0.000    &315.6315(2)                 &(315.61, 316.83)          &315.63       &315.630     &315.92 \\
0.005    &320.0105(2)                 &(319.99, 321.23)          &320.01       &320.009     &320.31  \\
0.010    &333.9323(2)                 &(333.91, 335.21)          &333.93       &333.931     &334.25  \\
0.015    &360.1322(2)                 &(360.10, 361.53)          &360.12       &360.130     &360.50  \\
0.020    &404.8286(2)                 &(404.79, 406.43)          &404.81       &404.825     &405.28  \\
0.025    &482.3411(3)                 &(482.29, 484.29)          &482.31       &482.335     &482.95   \\
0.030    &631.4758(3)                 &(631.39, 634.10)          &631.42       &631.463     &632.41    \\
0.035    &1001.751(2)                 &(1001.53, 1006.08)        &1001.59      &1001.71     &1003.68   \\
0.040    &3192.7(2)                   &(3190.16, 3207.18)        &3190.67      &3192.17     &3205.28    \\
0.045    &$-$2097.602(3)              &($-$14717.51, $-$2050.70) &$-$2098.66   &$-$2097.89  &$-$2097.0  \\
0.050    &$-$725.4477(2)              &($-$729.45, $-$718.73)    &$-$725.60    &$-$727.490  &$-$726.27  \\
0.055    &$-$416.4749(2)             &($-$419.07, $-$413.67)    &$-$416.54    &$-$416.492  &$-$417.20  \\
0.060    &$-$281.12719(2)            &($-$283.14, $-$279.41)    &$-$281.16    &$-$281.137  &$-$281.78   \\
0.065    &$-$205.59918(2)            &($-$207.31, $-$204.30)    &$-$205.63    &$-$205.606  &$-$206.21   \\
0.070    &$-$157.5702(2)              &($-$159.14, $-$156.44)    &$-$157.59    &$-$157.575  &$-$158.17   \\
0.075    &$-$124.33628(2)             &($-$125.87, $-$123.25)    &$-$124.35    &$-$124.340  &$-$124.94   \\
0.080    &$-$99.86344(2)              &($-$101.47, $-$98.72)     &$-$99.88     &$-$99.867   &$-$100.49   \\
0.085    &$-$80.8752(2)               &($-$82.67, $-$79.60)      &$-$80.89     &$-$80.878   &$-$81.55    \\
0.090    &$-$65.36352(2)             &($-$67.54, $-$63.84)      &$-$65.37     &$-$65.366   &$-$66.11    \\
0.095    &$-$51.88420(2)             &($-$54.81, $-$49.93)      &$-$51.92     &$-$51.888   &$-$52.74    \\
0.100    &$-$39.05227(2)             &($-$43.49, $-$36.31)      &$-$39.11     &$-$39.059   &$-$40.07   \\
0.105    &$-$24.69053(2)              &($-$32.65, $-$20.33)      &$-$24.82     &$-$24.709   &$-$26.03  \\
0.110    &$-$2.1515(2)                &                          &$-$2.66      &$-$2.248    &           \\
0.115    &93.381(2)                   &($-$6.27, 128.35)         &84.66        &91.175      &74.32     \\
0.120    &$-$125.6637(4)              &                          &$-$137.94    &            &           \\
\end{tabular}
\end{ruledtabular}
\end{table*}

Table~\ref{dynamic2} lists some selective values of dynamic dipole polarizabilities of
$2\,^3P$, $3\,^3S$, and $3\,^3P$ states for the He atom. For the $2\,^3P$ state,
we calculate the dynamic dipole polarizabilities for frequency below the
second excitation threshold, and for the $3\,^3S$ and $3\,^3P$ states, we only list the dynamic dipole polarizabilities
for frequency $\omega$ below the first ionization threshold. All of our results are very accurate
except few values for the frequency near resonance transition energy or ionization threshold.

\begin{figure}
\includegraphics[width=0.49\textwidth]{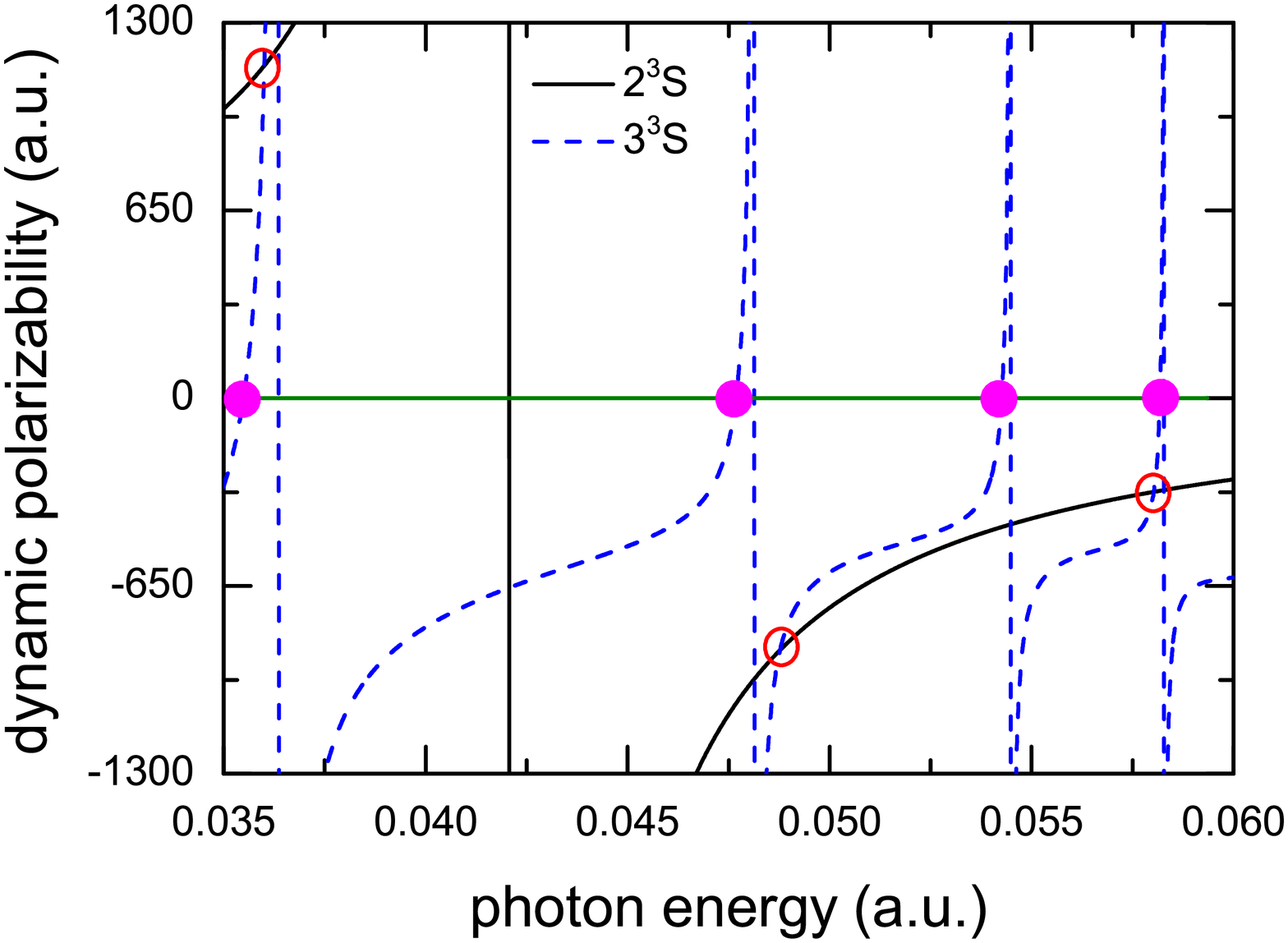}
\caption{\label{f1}(Color online) Dynamic dipole polarizabilities (in a.u.) of helium for the photon energy $0.035\leq\omega\leq 0.06$ a.u.
The solid black line denotes the dynamic polarizabilities for $2\,^3S$ state, and the dashed blue line represents the dynamic polarizabilities for $3\,^3S$ state. The crossing points denoted as solid magenta circle are the tune-out wavelengths, and the crossing points marked as blank red circle are the magic wavelengths. The vertical lines are the resonance transition positions, and the green line is a horizontal zero line. }
\end{figure}

\begin{figure}
\includegraphics[width=0.49\textwidth]{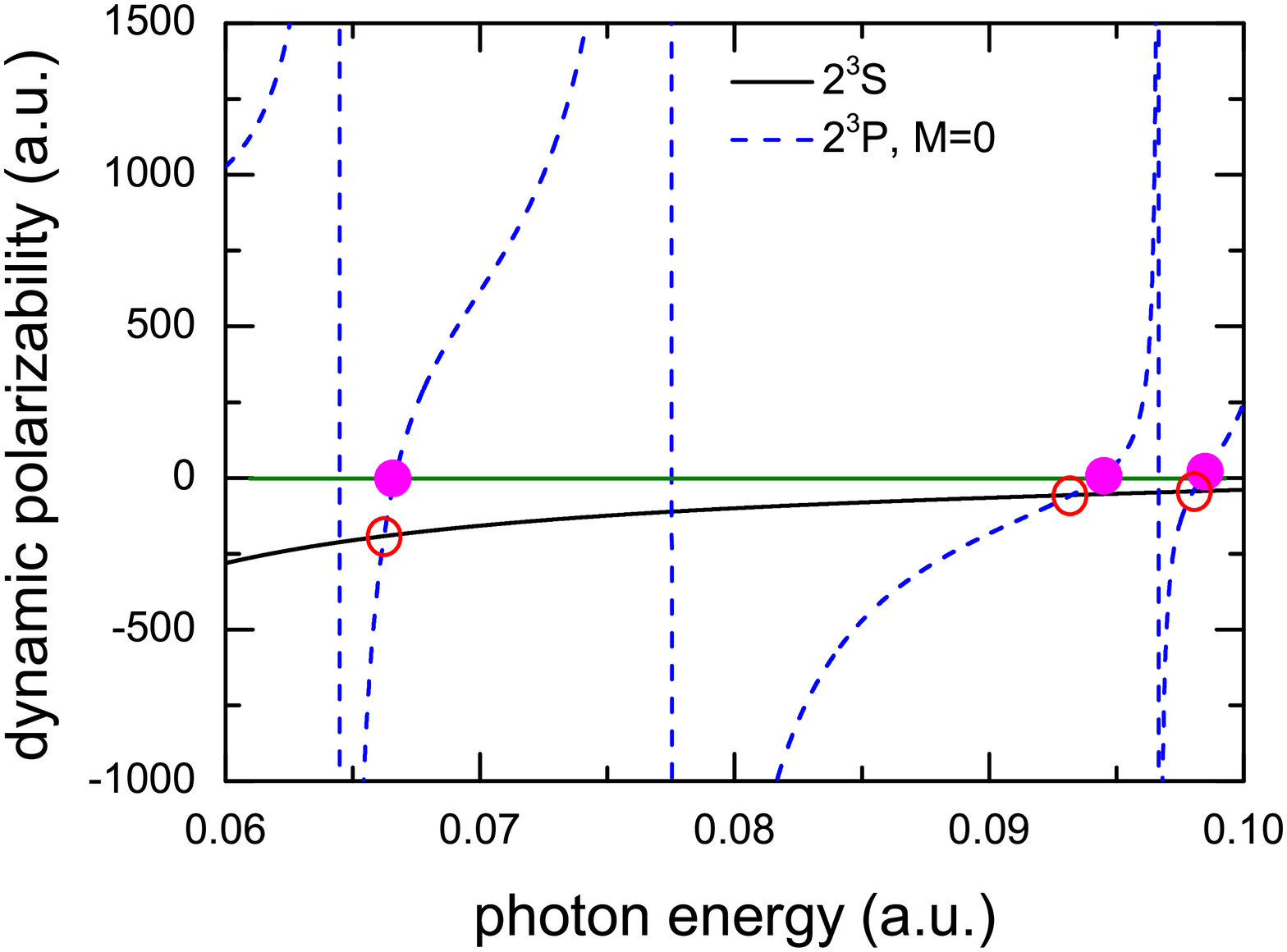}
\caption{\label{f2}(Color online) Dynamic dipole polarizabilities (in a.u.) of helium for the photon energy $0.06\leq\omega\leq0.10$ a.u.
The solid black line denotes the dynamic polarizabilities for $2\,^3S$ state, and the dashed blue line represents the dynamic polarizabilities for $2\,^3P (M=0)$ state.
The crossing points denoted as solid magenta circle are the tune-out wavelengths, and the crossing points marked as blank red circle are the magic wavelengths. The vertical lines are the resonance transition positions, and the green line is a horizontal zero line.}
\end{figure}

\begin{figure}
\includegraphics[width=0.49\textwidth]{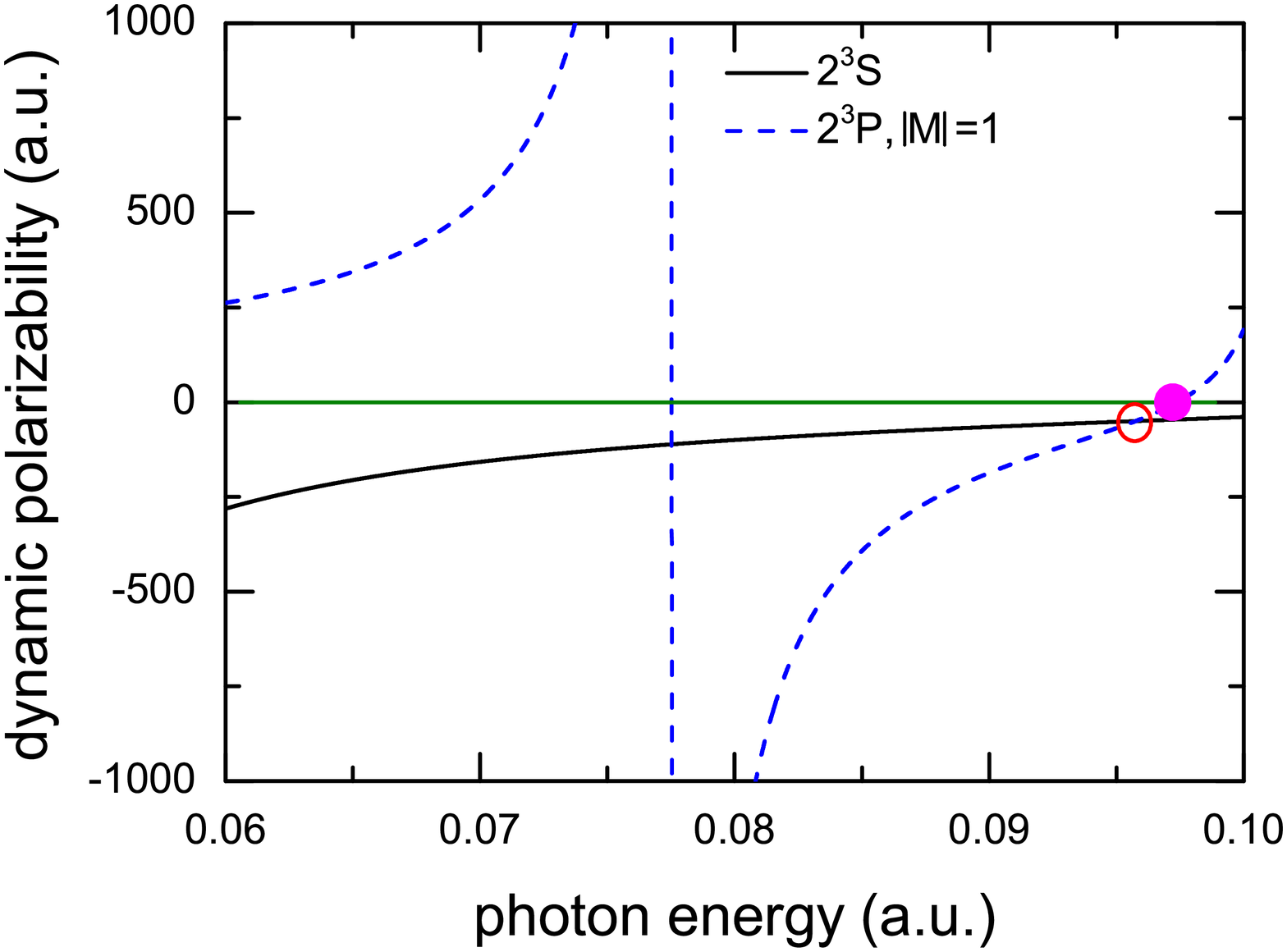}
\caption{\label{f3}(Color online) Dynamic dipole polarizabilities (in a.u.) of helium for the photon energy $0.06\leq\omega\leq0.10$ a.u. The solid black line denotes the dynamic polarizabilities for $2\,^3S$ state, and the dashed blue line represents the dynamic polarizabilities for $2\,^3P (|M|=1)$ state.
The crossing points denoted as solid magenta circle are the tune-out wavelengths, and the crossing points marked as blank red circle are the magic wavelengths. The vertical lines are the resonance transition positions, and the green line is a horizontal zero line.}
\end{figure}

\begin{figure}
\includegraphics[width=0.49\textwidth]{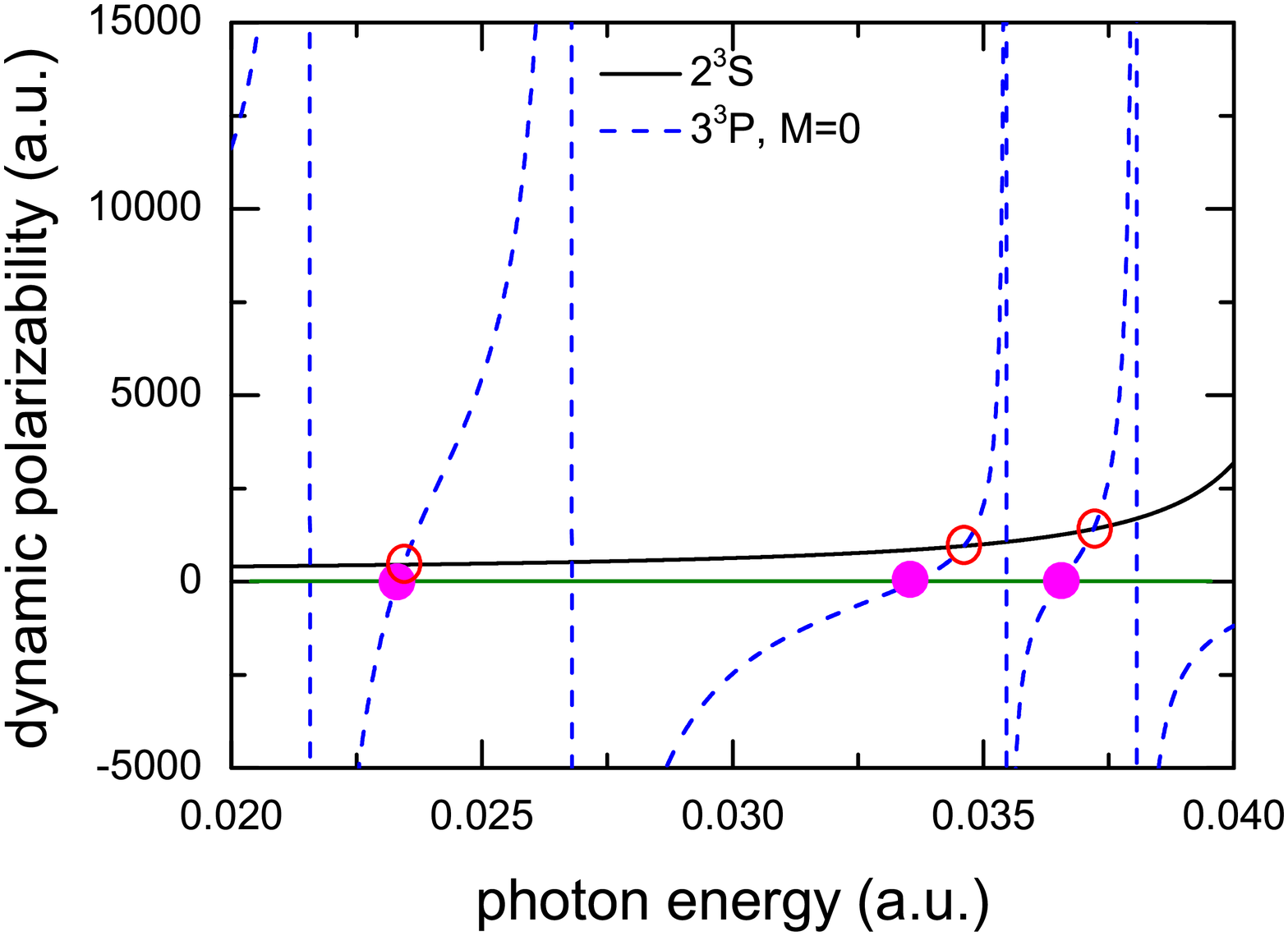}
\caption{\label{f4}(Color online) Dynamic dipole polarizabilities (in a.u.) of helium for the photon energy $0.02\leq\omega\leq0.04$ a.u.
The solid black line denotes the dynamic polarizabilities for $2\,^3S$ state, and the dashed blue line represents the dynamic polarizabilities for $3\,^3P (M=0)$ state.
The crossing points denoted as solid magenta circle are the tune-out wavelengths, and the crossing points marked as blank red circle are the magic wavelengths. The vertical lines are the resonance transition positions, and the green line is a horizontal zero line.}
\end{figure}

\begin{figure}
\includegraphics[width=0.49\textwidth]{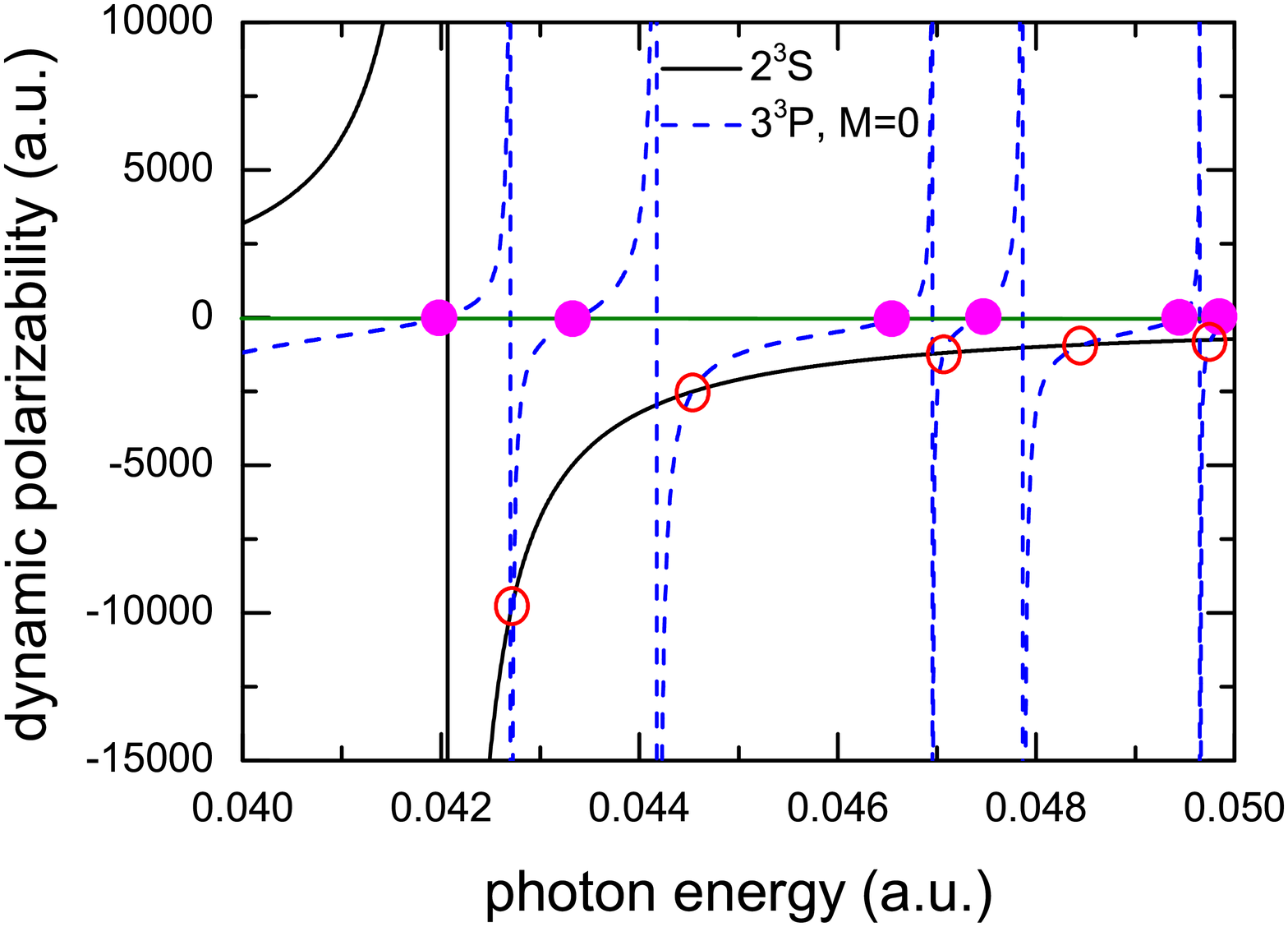}
\caption{\label{f5}(Color online) Dynamic dipole polarizabilities (in a.u.) of helium for the photon energy $0.04\leq\omega\leq0.05$ a.u.
The solid black line denotes the dynamic polarizabilities for $2\,^3S$ state, and the dashed blue line represents the dynamic polarizabilities for $3\,^3P (M=0)$ state.
The crossing points denoted as solid magenta circle are the tune-out wavelengths, and the crossing points marked as blank red circle are the magic wavelengths. The vertical lines are the resonance transition positions, and the green line is a horizontal zero line.}
\end{figure}

\begin{figure}
\includegraphics[width=0.49\textwidth]{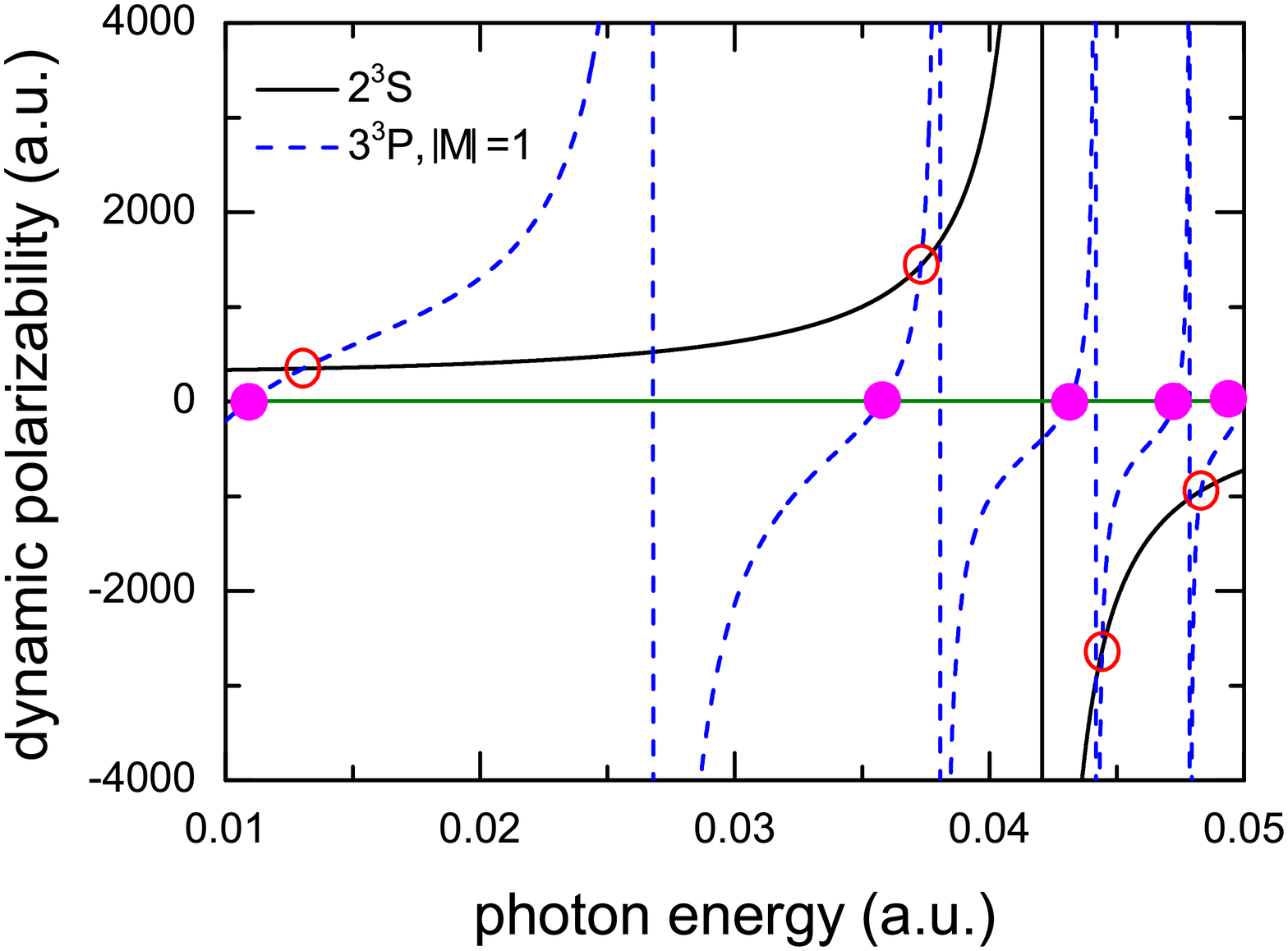}
\caption{\label{f6}(Color online) Dynamic dipole polarizabilities (in a.u.) of helium for the photon energy $0.01\leq\omega\leq0.05$ a.u.
The solid black line denotes the dynamic polarizabilities for $2\,^3S$ state, and the dashed blue line represents the dynamic polarizabilities for $3\,^3P (|M|=1)$ state.
The crossing points denoted as solid magenta circle are the tune-out wavelengths, and the crossing points marked as blank red circle are the magic wavelengths. The vertical lines are the resonance transition positions, and the green line is a horizontal zero line.}
\end{figure}

%=======================================================================================
\begin{table*}
\caption{\label{dynamic2}Dynamic polarizabilities (in a.u.) of $3\,^3S$, $2\,^3P$,
and $3\,^3P$ states for the He atom. The numbers in parentheses give the computational uncertainties.}
\begin{ruledtabular}
\begin{tabular}{ccccccc}
 \multicolumn{1}{c}{state}
&\multicolumn{1}{c}{$3\,^3S$}
&\multicolumn{2}{c}{$2\,^3P$}
&\multicolumn{2}{c}{$3\,^3P$} \\
\cline{1-1} \cline{2-2} \cline{3-4} \cline{5-6}
 \multicolumn{1}{c}{$\omega$}
&\multicolumn{1}{c}{$\alpha_1(\omega)$}
&\multicolumn{1}{c}{$\alpha_1(\omega)$} &\multicolumn{1}{c}{$\alpha_1^T(\omega)$}
&\multicolumn{1}{c}{$\alpha_1(\omega)$} &\multicolumn{1}{c}{$\alpha_1^T(\omega)$}\\
\hline
  0.000        &7937.584(2)        &46.70793(3)       &69.5964(2)       &17305.67(3)      &336.768(3)\\
  0.005        &10199.363(2)       &45.83274(4)       &70.8984(2)       &$-$8051.87(2)    &3534.464(2)\\
  0.010        &71124.62(5)        &42.97603(4)       &75.0559(2)       &$-$23498.47(2)   &23293.67(2)\\
  0.015        &$-$7892.735(2)     &37.3217(2)        &82.9492(2)       &3942.877(2)      &$-$3345.9627(4)\\
  0.020        &$-$3062.782(2)     &26.972(2)         &96.59195(3)      &4749.082(4)      &$-$3445.013(2)\\
  0.025        &$-$1685.264(2)     &7.4367(2)         &120.65874(3)     &4962.08(2)       &$-$246.35(2)\\
  0.030        &$-$1031.851(2)     &$-$33.77638(2)    &167.92335(6)     &$-$2249.18(2)    &111.731(2)\\
  0.035        &$-$305.1046(2)     &$-$145.8045(2)    &287.9533(2)      &502.52(2)        &$-$765.454(5)\\
  0.040        &$-$792.8559(2)     &$-$860.666(3)     &1013.44(2)       &$-$1078.384(2)   &53.0407(4)\\
  0.045        &$-$512.4321(2)     &924.3286(4)       &$-$757.2164(4)   &$-$1078.7(2)     &78.96(2)\\
  0.050        &$-$602.5701(5)     &498.7080(2)       &$-$311.6507(2)   &152(2)           &14.1(2)\\
  0.055        &$-$703.68(2)       &447.8331(3)       &$-$231.6643(2)   &                 &\\
  0.060        &$-$620.32(3)       &517.4623(6)       &$-$255.4211(4)   &                 &\\
  0.065        &                   &$-$563.39(2)      &907.7(2)         &                 &\\
  0.070        &                   &561.824(2)        &$-$27.2995(3)    &                 &\\
  0.075        &                   &1635.892(8)       &$-$167.107(2)    &                 &\\
  0.080        &                   &$-$1497.409(7)    &137.702(2)       &                 &\\
  0.085        &                   &$-$418.409(2)     &26.1352(2)       &                 &\\
  0.090        &                   &$-$185.5369(3)    &$-$1.0091(2)     &                 &\\
  0.095        &                   &$-$29.424(2)      &$-$39.1638(3)    &                 &\\
  0.100        &                   &211.866(3)        &$-$18.9742(3)    &                 &\\
\end{tabular}
\end{ruledtabular}
\end{table*}
The dynamic dipole polarizabilities for the lowest four triplet states of helium are also plotted in the Figs.~\ref{f1}-~\ref{f6}
as the photon energy $\omega$. For the non-zero angular momentum state, the polarizability depends upon its magnetic quantum
number $M$ because of both scalar and tensor polarizabilities existing, so the dynamic dipole polarizabilities for $2\,^3P$ and $3\,^3P$ states
are divided into two cases as $M=0$ and $|M|=1$. The crossing points between a curve and the horizontal zero line are called as tune-out wavelengths, which denoted as solid magenta circle, and the crossing points between two curves are the magic wavelengths, which denoted as blank red circle. The vertical lines are the resonance transition positions.

\subsection{Tune-Out Wavelengths}

%=======================================================================================
\begin{table*}
\caption{\label{tuneout}Tune-out wavelengths ($\lambda_t$) for $2\,^3S$, $3\,^3S$, $2\,^3P$, and $3\,^3P$ states of helium.
The second and third column are the tune-out wavelengths in atomic units and nanometer respectively.
The numbers in parentheses give the computational uncertainties. }
\begin{ruledtabular}
\begin{tabular}{lll}
 \multicolumn{1}{l}{State}
&\multicolumn{1}{c}{$\omega_{t}$(a.u.)}             &\multicolumn{1}{c}{$\lambda_{t}$(nm)}
 \\
\hline
$2\,^3S$  &{\bf 0.110 312 66(2)} &{\bf413.038 28(3)} \\
\hline
$3\,^3S$  &0.035 488 102(2)     &1283.905 03(4)\\
          &0.047 681 245(3)     &955.582 28(7)\\
          &0.054 260 56(3)      &839.714(2) \\
          &0.058 1756(2)        &783.204(2) \\
\hline
$2\,^3P(M=0)$ &0.066 652 71(2)  &683.5934(2)\\
              &0.094 341 03(2)  &482.9644(2) \\
              &0.098 338 54(2)  &463.3316(2) \\
\hline
$2\,^3P(M=\pm1)$ &0.097 382 82(2)  &467.8788(2) \\
\hline
$3\,^3P(M=0)$ &0.023 315 997(3)    &1954.1670(3) \\
              &0.033 637 088(2)    &1354.5570(2) \\
              &0.036 548 517(2)    &1246.6539(3) \\
              &0.042 005 02(2)     &1084.712(2) \\
              &0.043 390 63(3)     &1050.074(2)\\
              &0.046 5939(2)       &977.883(2) \\
              &0.047 4031(2)       &961.189(3)  \\
              &0.049 4427(2)       &921.539(4) \\
              &0.049 9644(2)       &911.917(4) \\
\hline
$3\,^3P(M=\pm1)$ &0.010 911 33(2)  &4175.783(4)  \\
                 &0.035 787 67(2)  &1273.1580(2)\\
                 &0.043 149 27(2)  &1055.9472(3) \\
                 &0.047 2945(1)    &963.397(2) \\
                 &0.049 9054(1)    &912.995(3) \\
\end{tabular}
\end{ruledtabular}
\end{table*}
Table~\ref{tuneout} lists the values of tune-out wavelengths in the 400-4200 nm region, which marked as solid magenta circle in the Figs.~\ref{f1}-~\ref{f6},
For the metastable state of helium, Mitroy and Tang~\cite{mitroy13b} have obtained the 413.02(9)~nm tune-out wavelength
by incorporating Hylleraas matrix elements for the transition to $2\,^3P$ and $3\,^3P$ manifolds and core-polarization model
matrix elements for other transitions, and they predicted the tune-out wavelength around 413~nm can be used to test
the QED effect. Recently, a experimental measurement of Ken Baldwin's group report the tune-out wavelength being 413.0938(9Stat. )(20Syst.) nm~\cite{henson15}
and another theoretical calculation by Notermans~{\em et al}~\cite{notermans14} gives 414.197~nm tune-out wavelength
by using available tables of level energies and Einstein A coefficients. Our tune-out wavelength of ab-initio calculation is 413.038 28(3)~nm,
which corroborates the value 413.02(9) nm of Mitroy and Tang~\cite{mitroy13b}.
The difference between the theoretical calculations and the experimental measurement may caused by finite nuclear mass,
relativistic and QED corrections, which calls for great efforts for theoretical calculation to improve the precision for QED test.

\subsection{ Magic Wavelengths}

%=======================================================================================
\begin{table*}
\caption{\label{magic} Magic wavelengths ($\lambda_m$) for $2\,^3S \to 3\,^3S$, $2\,^3S \to 2\,^3P(M=0, \pm1)$
and $2\,^3S \to 3\,^3P(M=0, \pm 1)$ transitions of the He atom. The corresponding dynamic
dipole polarizabilities at the magic wavelengths are given in the last column.
The numbers in parentheses give the computational uncertainties.}
\begin{ruledtabular}
\begin{tabular}{llll}
\multicolumn{1}{l}{Transition}
&\multicolumn{1}{c}{$\omega_{m}(a.u.)$}
&\multicolumn{1}{c}{$\lambda_{m}(nm)$}
&\multicolumn{1}{c}{$\alpha_1(\omega_{m})(a.u.)$}\\
\hline
$2\,^3S \to 3\,^3S$     &0.036 004 592(2)     &1265.48724(4)    &1151.058(2)\\
                        &0.048 775 162(5)     &934.1507(2)      &$-$872.007(2)\\
                        &0.058 0244(2)        &785.245(3)       &$-$324.31(2)\\
\hline
$2\,^3S \to 2\,^3P(M=0)$  &0.066 228 54(2)    &687.9716(2)     &$-$191.9157(2)\\
                          &0.093 2295(2)      &488.7225(2)     &$-$56.5068(2)\\
                          &0.098 0166(2)      &464.8532(3)     &$-$44.169 53(4)\\
\hline
$2\,^3S \to 2\,^3P(M=\pm1)$ &0.095 7473(2)   &475.870 85(6)    &$-$49.962 90(4)\\

\hline
$2\,^3S \to 3\,^3P(M=0)$  &0.023 434 415(3)       &1944.2923(3)      &453.0297(2)\\
                          &0.034 618 094(3)       &1316.1716(2)      &955.639(2)\\
                          &0.037 202 6377(2)      &1224.73446(2)     &1410.365(2)\\
                          &{\bf 0.042 734 44(3)}  &{\bf 1066.197(2)} &$-$9487(2)\\
                          &0.044 532 37(2)        &1023.151(2)       &$-$2511.07(3)\\
                          &0.047 069 3(2)         &968.007(3)        &$-$1196.32(3)\\
                          &0.048 478 5(2)         &939.867(4)        &$-$915.97(3)\\
                          &0.049 762 5(3)         &915.616(5)        &$-$750.19(3)\\
\hline
$2\,^3S \to 3\,^3P(M=\pm1)$  &0.013 031 86(2)     &3496.304(3)      &348.0648(2)\\
                             &0.037 297 86(2)     &1221.607 75(2)   &1436.592(2)\\
                             &0.044 421 15(2)     &1025.713(2)      &$-$2633.59(3)\\
                             &0.048 309 5(2)      &943.156(4)       &$-$942.89(2)\\
\end{tabular}
\end{ruledtabular}
\end{table*}
The magic wavelength is the wavelength at which the polarizability difference for a
transition goes to zero, which means the first-order Stark shifts for the upper and lower levels
of a transition are the same~\cite{ye99, safronova13g}.
Table~\ref{magic} presents all the values of magic wavelengths in the 460-3500 nm region marked in blank red circle in the Figs.~\ref{f1}-\ref{f6}.
The corresponding dynamic dipole polarizabilities at the magic wavelengths are also given in the last column.
For the magic wavelength of 1066.197(2) nm, there exist two terms, which play
major contribution of the dynamic dipole polarizabilities for the $2\,^3S$ and $3\,^3P(M=0)$ states respectively.
Table~\ref{analyse} lists some contributions from different intermediate states for the $2\,^3S \to 3\,^3P(M=0)$ transition in detail
at the magic wavelength of 1066.197(2) nm. We can see that the contribution from $2\,^3P$ state to the polairzability of $2\,^3S$ is about 99.87\%, and
the contribution from $6\,^3S$ state to the polairzability of $3\,^3P(M=0)$ is about 98.87\%. According to the definition of magic wavelength $\alpha_{2\,^3S}(\omega_m)=\alpha_{3\,^3P}(\omega_m)$, we have the expanded form,
\begin{eqnarray}
\frac{f_{2\,^3S \to 2\,^3P}}{\Delta E_{2\,^3S \to 2\,^3P}^2-\omega_m^2}+\alpha_{2\,^3S}(Remainder;\omega_m)&=& \frac{3f_{3\,^3P \to 6\,^3S}}{\Delta E_{3\,^3P \to 6\,^3S}^2-\omega_m^2}+\alpha_{3\,^3P}(Remainder;\omega_m) \nonumber \label{eq25}\\
\end{eqnarray}
where the second term in the left of Eq.~(\ref{eq25}) is all the contributions from other $n\,^3P$ states to the dynamic dipole polarizability of $2\,^3S$ state,
and the second term in the right of Eq.~(\ref{eq25}) is all the contributions from other $n\,^3S$, $n\,^3D$, and $n\,^3P^e$ states to the dynamic dipole polarizability of $3\,^3P (M=0)$ state. If all the remainder terms are neglected, then the ratios of oscillator strengths and reduced matrix elements are written as
\begin{eqnarray}
\frac{f_{2\,^3S \to 2\,^3P}}{f_{3\,^3P \to 6\,^3S}}&=& \frac{3(\Delta E_{2\,^3S \to 2\,^3P}^2-\omega_m^2)}{\Delta E_{3\,^3P \to 6\,^3S}^2-\omega_m^2}  \label{eq26}\\
\frac{M_{2\,^3S \to 2\,^3P}}{M_{3\,^3P \to 6\,^3S}}&=& \frac{|\langle 2\,^3S\|
\sum\limits_{i=1, 2}r_iC^{(1)}(\hat{r}_i)
                                \|2\,^3P\rangle|}{|\langle 3\,^3P\|
\sum\limits_{i=1, 2}r_iC^{(1)}(\hat{r}_i)
                                \|6\,^3S\rangle|}= \sqrt{\frac{\Delta E_{3\,^3P \to 6\,^3S}(\Delta E_{2\,^3S \to 2\,^3P}^2-\omega_m^2)}{\Delta E_{2\,^3S \to 2\,^3P}(\Delta E_{3\,^3P \to 6\,^3S}^2-\omega_m^2)}}  \label{eq27}
\end{eqnarray}
Combined present energy difference and the magic wavelength 1066.197(2) nm, the ratios of the oscillator strengths and the reduced matrix elements are determined and listed in Table~\ref{xi}. Present$^1$ are the values of our ab-initio calculation, and Present$^2$ are derived by substituting our theoretical energies and the magic wavelength of 1066.197(2) nm into the Eqs.(\ref{eq26}) and (\ref{eq27}). Compared with the explicitly correlated results of Ref.~\cite{cann92}, we believe our values of Present$^1$ are reliable, since present oscillator strengths for $2\,^3S \to 2\,^3P$ and $3\,^3P \to 6\,^3S$ transitions are much more accurate than the values of Ref.~\cite{cann92} by at least one order of magnitude. In order to test the accuracy of the values derived from Eqs.(\ref{eq26}) and (\ref{eq27}), we can compare the results between Present$^1$ and Present$^2$. It's clearly seen that
the derived values 64.6653 and 4.677847 from the Eqs.(\ref{eq26}) and (\ref{eq27}) are in good agreement with our ab-initio values 65.48(2) and 4.7073(3) at the level of 1.3\%  and 0.7\% accuracy respectively. If increasing the number of B-spline basis sets, and also considered the contribution of the remainder term, then improvement of the accuracy for the transition matrix elements ratio $(M_{2\,^3S \to 2\,^3P})/(M_{3\,^3P \to 6\,^3S})$ up to 0.5\% is achievable.

As we known that, present experimental technique is very difficult to measure matrix elements accurately, only 1\% accuracy for one or two of
the lowest transitions have been reported~\cite{gomez04,bouloufa09a}. Recently, Herold~{\em et al.} present a method for accurate determination of $5s-6 p$ matrix elements in rubidium by measurements of the ac Stark shift around tune-out wavelength~\cite{herold12a}. In our calculation, the particular magic wavelength around 1066 nm can be used for experiment measurement to determine the atomic transition matrix elements involved highly excited states for helium.

%=============================================================================================
\begin{table*}
\caption{\label{analyse}Contributions from some of intermediate states to the dynamic dipole polarizability of $2\,^3S$ and $3\,^3P(M=0)$ states at the magic wavelengths 1066.197(2) nm.}
\begin{ruledtabular}
\begin{tabular}{lc}
\multicolumn{1}{l}{$\omega(a.u.)$} &\multicolumn{1}{c}{0.042 734 44(3)}
\\
\multicolumn{1}{l}{$\lambda(nm)$} &\multicolumn{1}{c}{1066.197(2)}
\\
 \hline
\multicolumn{1}{c}{Intermediate states} &\multicolumn{1}{c}{$2\,^3S$}\\
$2\,^3P$  &{\bf $-$9499.066}  \\
$3\,^3P$  &5.418        \\
$4\,^3P$  &1.386        \\
Others &5.234        \\
Total     &{\bf $-$9487.028}  \\
 \hline
\multicolumn{1}{c}{Intermediate states} &\multicolumn{1}{c}{$3\,^3P(M=0)$} \\
$3\,^3S$  &519.840         \\
$4\,^3S$  &$-$320.109    \\
$5\,^3S$  &$-$118.888     \\
$6\,^3S$  &{\bf $-$9379.600}    \\
$3\,^3D$  &$-$73.9017   \\
$4\,^3D$  &$-$517.059   \\
$5\,^3D$  &$-$395.758   \\
$6\,^3D$  &506.415      \\
Others    &292.0327     \\
Total     &{\bf $-$9487.028}  \\
\end{tabular}
\end{ruledtabular}
\end{table*}

\begin{table*}
\caption{\label{xi}Comparison of the ratios for the oscillator strengths $(f_{2\,^3S \to 2\,^3P})/(f_{3\,^3P \to 6\,^3S})$ and the reduced matrix elements $(M_{2\,^3S\to 2\,^3P})/(M_{3\,^3P\to 6\,^3S})$. Present$^1$ are the value of our ab-initio calculation, and Present$^2$ are derived by substituting our theoretical energies and the magic wavelength of 1066.197(2) nm into the Eqs.(\ref{eq26}) and (\ref{eq27}). The numbers in parentheses give the computational uncertainties.}
\begin{ruledtabular}
\begin{tabular}{lcc}
 \multicolumn{1}{c}{}
&\multicolumn{1}{c}{$(f_{2\,^3S \to 2\,^3P})/(f_{3\,^3P \to 6\,^3S})$ }  &\multicolumn{1}{c}{$(M_{2\,^3S\to 2\,^3P})/(M_{3\,^3P\to 6\,^3S})$ }\\
\hline
Present$^1$          &65.48(2)    &4.7073(3) \\
Present$^2$          &64.6653     &4.677847 \\
Ref.~\cite{cann92}   &65.5308     &4.789739 \\
\end{tabular}
\end{ruledtabular}
\end{table*}

\section{conclusions}

The calculations of the energies and the main oscillator strengths for the four triplet
states ($2\,^3S$, $3\,^3S$, $2\,^3P$, and $3\,^3P$) in the length, velocity and
acceleration gauges are carried out by the configuration interaction based on the B-spline functions.
Also the accurate dynamic dipole polarizabilities for the four lowest triplet
states are obtained. Ours static dipole polarizabilities in the length and velocity gauges have 5-6 significant digits, which are in excellent
agreement with the variational Hylleraas calculations. Present work lays solid foundation for the
further to calculate the relativistic and QED effects on the dynamic polarizabilities of helium.

In particular, the tune-out wavelengths for the four triplet states and magic wavelengths for the three transitions of $2\,^3S \to 3\,^3S$, $2\,^3S \to 2\,^3P$, and
$2\,^3S \to 3\,^3P$ are determined with high precision. Our tune-out wavelength 413.038 28(3) nm of the metastable state validate the value of Mitroy and Tang~\cite{mitroy13b}. And the magic wavelength around 1066 nm for $2\,^3S \to 3\,^3P$ transition is proposed for experimental measurement
to determine the ratio of the transition matrix elements ($2\,^3S \to 2\,^3P)/(3\,^3P \to 6\,^3S$), this is
a unique way to obtain accurate transition matrix element involved highly excited states.
Also we expected that other tune-out wavelengths and magic wavelengths can provide theoretical reference for the precision-measurement experiment design in the future.

\begin{acknowledgements}

This work was supported by NNSF of China under Grant Nos. 11474319, 11274348, and by the National Basic Research Program of China under
Grant No. 2012CB821305. This work is dedicated to Professor James Mitroy of Charles Darwin
University, who unexpectedly passed away shortly after suggestion of this work.

\end{acknowledgements}

%\bibliography{positron}

\end{document}